\documentclass[twocolumn,showpacs,superscriptaddress,preprintnumbers,nofootinbib,prd]{revtex4-1}
\usepackage{amsmath,amssymb,amsfonts,slashed}
\usepackage{graphicx}
\usepackage{graphics}
\usepackage{dcolumn}
\usepackage{epsfig,color}
\usepackage{epstopdf}
\usepackage{bm}
\usepackage{verbatim}
\usepackage{epstopdf}
\def\be{\begin{equation}}
\def\ee{\end{equation}}

\newcommand{\blue}[1]{{\color{blue}#1}}

\begin{document}

\title{Spectroscopy and decays of the fully-heavy tetraquarks}

\author{M. Naeem Anwar}\email[]{naeem@itp.ac.cn}
\affiliation{CAS Key Laboratory of Theoretical Physics, Institute of Theoretical Physics, Chinese Academy of Sciences, Beijing 100190, China}
\affiliation{University of Chinese Academy of Sciences, Beijing 100049, China}
\affiliation{Helmholtz-Institut f\"ur Strahlen- und Kernphysik and Bethe
Center for Theoretical Physics, \\Universit\"at Bonn,  D-53115 Bonn, Germany}

\author{J. Ferretti}
\affiliation{CAS Key Laboratory of Theoretical Physics, Institute of Theoretical Physics, Chinese Academy of Sciences, Beijing 100190, China}

\author{F.-K. Guo}
\affiliation{CAS Key Laboratory of Theoretical Physics, Institute of Theoretical Physics, Chinese Academy of Sciences, Beijing 100190, China}
\affiliation{School of Physical Sciences, University of Chinese Academy of
Sciences, Beijing 100049, China}

\author{E. Santopinto}
\affiliation{INFN, Sezione di Genova, via Dodecaneso 33, 16146 Genova, Italy}

\author{B.-S. Zou}
\affiliation{CAS Key Laboratory of Theoretical Physics, Institute of Theoretical Physics, Chinese Academy of Sciences, Beijing 100190, China}
\affiliation{School of Physical Sciences, University of Chinese Academy of
Sciences, Beijing 100049, China}

\date{\today}

\begin{abstract}

  We discuss the possible existence of the fully-heavy tetraquarks. We calculate
  the ground-state energy of the $bb \bar b \bar b$ bound state, where $b$
  stands for the bottom quark, in a nonrelativistic effective field theory
  framework with one-gluon-exchange (OGE) color Coulomb interaction, and in a
  relativized diquark model characterized by OGE plus a confining potential.
  Our analysis advocates the existence of uni-flavor heavy four-quark bound
  states. The ground state $bb\bar b\bar b$ tetraquark mass is predicted to be
  $(18.72\pm0.02)$~GeV. Mass inequality relations among the lowest
  $QQ\bar{Q}\bar{Q}$ state, where $Q\in \{c, b\}$, and the corresponding heavy
  quarkonia are presented, which give the upper limit on the mass of ground state
  $QQ\bar{Q}\bar{Q}$. The possible decays of the lowest $bb\bar{b}\bar{b}$
  are highlighted, which might provide useful references in the search for them in
  ongoing LHC experiments, and its width is estimated to be a few tens of MeV.

\end{abstract}


\maketitle

\section{Introduction}

Heavy quarkonium spectroscopy is the arena of quark potential models, in which
the full interaction incorporates the short-range one-gluon-exchange (OGE) and
long-range confining
potential~\cite{Eichten:1978tg,Richardson:1978bt,Buchmuller:1980su,Godfrey:1985xj,Barnes:2005pb}.
To put heavy quarkonium spectroscopy in a model-independent framework, efforts have
been made to develop a nonrelativistic effective field theory
formalism~\cite{Pineda:1997bj,Brambilla:1999xf,Brambilla:2004jw}.

Higher-order perturbative calculations show that the lowest-lying heavy
quarkonia can be assigned as weakly coupled
states~\cite{Pineda:1997hz,Brambilla:1999xj,Kniehl:2002br}. They are
characterized by a momentum scale $mv \gg \Lambda_{\rm QCD}$, where $m$ and $v$
are the heavy quark mass and velocity, respectively, and $\Lambda_{\rm QCD}$ is the
nonperturbative scale of quantum chromodynamics (QCD), and their dynamics is
mainly dominated by the short-distance interaction. In particular, there is
some evidence that the first few bottomonia and $B_c$ mesons can be identified
as weakly coupled
systems~\cite{Titard:1993nn,Brambilla:2000db,Brambilla:2001fw}, while the
$J/\psi$ lies on the borderline of this classification~\cite{Brambilla:2004jw}.
The lowest-lying doubly- and triply-heavy baryons have also been studied as weakly
coupled bound states~\cite{Brambilla:2005yk,Jia:2006gw} but, due to the lack of
experimental incentive,\footnote{Very recently, the LHCb Collaboration announced the first
measurement of the $S$-wave doubly-charm baryon, the
$\Xi_{cc}^{++}$~\cite{Aaij:2017ueg}, with a mass of about 3620~MeV.}
this area is not as well explored as that of heavy
quarkonia.

It is also worth mentioning some recent indications of the existence of
mesons whose properties and quantum numbers do not fit into a traditional
quarkonium interpretation (for recent reviews,
see~\cite{Nakamura:2010zzi,Brambilla:2014jmp,Chen:2016qju,Lebed:2016hpi,
Esposito:2016noz,Guo:2017jvc,Ali:2017jda,Olsen:2017bmm}).
These states, including $Z_b(10610)$ and $Z_b(10650)$~\cite{Bondar},
$Z_c(3900)$~\cite{Ablikim,Liu}, and $Z_c(4025)$~\cite{Ablikim2},  have similar
features and must be made up of four valence quarks, as they are isovector
states in the heavy quarkonium mass region.
Another interesting example is $X(3872)$~\cite{Choi:2003ue} which, because
of its unusual properties, does not fit well into a pure charmonium
picture~\cite{Godfrey:1985xj,Eichten:1978tg,Barnes:2005pb}.
Several alternative interpretations have been
proposed~\cite{Hanhart:2007yq,Baru:2011rs,Swanson:2003tb,Pennington:2007xr,
Li:2009ad,Danilkin:2010cc,charmonium,Maiani:2004vq}.

Fully-heavy tetraquarks, such as $bb\bar{b}\bar{b}$ and the charm analog
$cc\bar{c}\bar{c}$, are of considerable interest, since they are free of light
degrees of freedom and might be used as an elegant probe to investigate the interplay
between both aspects of QCD: perturbative and nonperturbative. Owing to the lack
of experimental incentive, only a few theoretical studies have been conducted in this
field, and whether Nature allows the bound states of fully-heavy tetraquarks or
not is still an open question.
If it does, what are their masses and decay properties? Recently, this debate has given rise to
several studies within different approaches, including the constituent quark and diquark model~\cite{Bai:2016int,Eichten:2017ual,Karliner:2016zzc,Berezhnoy:2011xn},
chiral quark model~\cite{Wu:2016vtq}, chromo-electric potential model~\cite{Richard:2017vry}, and QCD sum
rules~\cite{Chen:2016jxd,Wang:2017jtz}. All of these predict the existence of
fully-heavy (bottom) tetraquarks, except for Ref.~\cite{Richard:2017vry}, in which
the authors argue that the stability of a fully-heavy system of quarks should rely on more subtle effects
that are not included in the simple picture of constituent quarks.
The fully-heavy tetraquark case is similar to that of polyelectrons
$\textrm{Ps}_{2}$, the bound state of two electrons and two positrons ($e^- e^- e^+ e^+$), discussed
long ago by Wheeler~\cite{Wheeler:1946zz}. Just after the prediction of $\textrm{Ps}_{2}$, a debate on their stability started
\cite{Ore:1946zz,Hylleraas:1947zza}, but it took half a century to get experimental
evidence of them~\cite{Nature:2007}.

In this paper, we compute the $b b \bar b \bar b$ ground-state energy. The
calculations are performed within the formalism of a nonrelativistic effective
field theory (NREFT) at the leading order (LO), which neglects all
spin-dependent and confining interactions, and in a relativized
diquark-antidiquark model, both of which are characterized by the OGE potential.
We also make a rough estimate of the $b b \bar b \bar b$ total width, in order to guide experimental
searches.

The paper is organized as follows. In section~\ref{nrmodel}, we describe the
NREFT approach to solve the system considered.
At the end of this section, we include our diquark model results.
Section~\ref{resultDiss} is devoted to discussing the results, and we compare
our predictions of the mass of the $X_{bb\bar{b}\bar{b}}$ ground-state with those
of previous studies. In Section~\ref{massInqSec}, we derive a set of mass
inequalities, and provide the upper limits of the ground state masses of several
fully-heavy tetraquarks.
In Sec.~\ref{decays}, we highlight some possible decay
modes of the $X_{bb\bar{b}\bar{b}}$ and give the ballpark estimate of its total
decay width. Finally, we provide a short summary.

\section{Formalism}
\label{nrmodel}

In this section, we describe an NREFT approach to the fully-heavy tetraquark
spectroscopy, characterized by the OGE interaction. This is used to estimate the
mass of the $b b \bar b \bar b$ ground state.
As all of the four quarks are very heavy, the tetraquark system is weakly
coupled with a small size of the order $1/(m_Q\alpha_s)\sim 1/(m_Q v)$,
where $v\ll1$ is the heavy quark velocity.
In this case, the dynamics of the system is dominated by the OGE which provides
a color Coulomb potential at the LO, and the long-distance confining
potential and spin-dependent interactions become perturbations.

In addition, we also give a tetraquark (diquark-antidiquark) model
prediction for the $b b \bar b \bar b$ ground-state mass.

\subsection{Nonrelativistic Hamiltonian}

The nonrelativistic Hamiltonian describing the $2Q-2\bar{Q}$ system
takes the following form at the LO
\begin{equation}
	\label{eqn:nonrelH}
	\mathcal H^{\rm NR}=\sum_{i=1}^{4} T_i + \sum_{i<j}
	V_{\textrm{SI}}(\textbf{r}_{ij}) \mbox{ },
\end{equation}
where $T_i = m_i +{\textbf{p}_{i}^{2}}/{(2m_i)}$,
$\textbf{r}_{ij}\equiv|\textbf{r}_{i}-\textbf{r}_{j}|$,
and the second term is the spin-independent OGE color Coulomb potential,
\begin{equation}
V_{\textrm{SI}}(\textbf{r}_{ij})=\sum_{i<j} \frac{\bm\lambda_i}{2} \cdot  \frac{\bm\lambda_j}{2} \frac{\alpha_s}{|\textbf{r}_{i}-\textbf{r}_{j}|} \mbox{ },
\end{equation}
where $\bm\lambda_i$ and $\bm\lambda_j$ are color matrices and $\alpha_s$ is the strong coupling strength.
The quark mass $m_b$ is to be determined from the ground state
bottomonium masses, and $\alpha_s$ is taken at the scale of the
typical momentum transfer. At order $\alpha_{s}^{2}$, we can write $m_b$ as~\cite{Jia:2006gw}
\begin{equation}
  \label{eqn:bmass}
  m_b = \frac{M_{b\bar b(1S)}}{2}\left( 1+ \frac29 \alpha_s^2(\mu) \right).
\end{equation}
Here, we take the spin-averaged mass $M_{b\bar b(1S)}=\left(M_{\eta_{b}}+3
M_{\Upsilon} \right)/4 $.

\subsection{Solving the four-body problem}

The four-body problem is notoriously delicate. One should be careful about the
choice of the wave functions, because a crude adoption may give rise to
misleading conclusions, as first illustrated by Ore in 1946 in Ref.
\cite{Ore:1946zz} in the case of polyelectrons ($e^- e^- e^+ e^+$ bound states).
One year later the author, in collaboration with
Hylleraas~\cite{Hylleraas:1947zza}, came up with an elegant prescription to
handle four-body systems.
This is what we are willing to use to calculate the $bb \bar b \bar b$
ground-state energy, provided that we make the substitution $e^- \rightarrow b$,
$e^+ \rightarrow \bar b$.

\begin{figure}[h]
\begin{center}
\includegraphics[width=5.5cm]{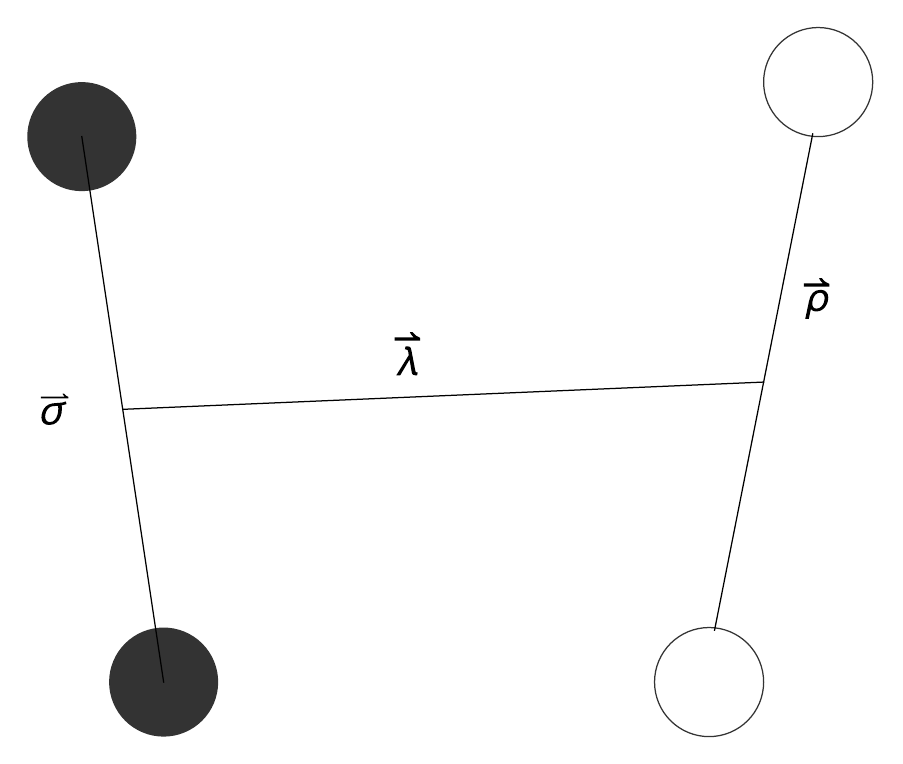}
\end{center}
\caption{Tetraquark Jacobi coordinates of Eq. (\ref{eqn:Jacobi}). Filled and
open circles represent quarks and antiquarks, respectively.}
\label{config}
\end{figure}

Since the ground state is non-degenerate and its symmetries are governed by
those of the Hamiltonian $\mathcal H^{\rm NR}$, the spatial wave function should
be symmetric under the exchange of $Q_1Q_2\leftrightarrow \bar{Q}_3 \bar{Q}_4$.
This leads to $\psi_{\textrm{spatial}}(Q_1Q_2)=\psi_{\textrm{spatial}}(\bar{Q}_3
\bar{Q}_4)$. This symmetry~\cite{Hylleraas:1947zza} helps to reduce the
number of integration variables and simplify the four-body problem.
To describe the quark relative motion, we define the following Jacobi
coordinates,
\be
\label{eqn:Jacobi}
{\bm\sigma} = {\bf r}_1 - {\bf r}_2 ~, ~~ {\bm\rho} = {\bf r}_3 - {\bf r}_4 ~, ~~ {\bm \lambda} = \frac{1}{2} ({\bf r}_1 + {\bf r}_2 - {\bf r}_3 - {\bf r}_4) ~,
\ee
which are shown in Fig.~\ref{config}.

The ``physical" tetraquark color wave function can be written as
a superposition of different color configurations,\footnote{In the following,
we use the notation $\left| \bf{C}_{12} \bf{C}_{34}; \bf{C}_{1234} \right\rangle$,
where the color configuration of the quark (antiquark) pairs 12 and 34,
$\bf{C}_{12}$ and $\bf{C}_{34}$, are coupled to that of the tetraquark, $\bf{C}_{1234}$.}
\begin{equation}
	\label{eqn:Psi-c-tetra-phys}
	\left|\psi_{\rm c} \right\rangle = \alpha \left| \bar{\textbf{3}}_{12}\textbf{3}_{34};
    \textbf{1}_{1234} \right\rangle + \beta \left| \textbf{6}_{12}\bar{\textbf{6}}_{34};
    \textbf{1}_{1234} \right\rangle
	\mbox{ },
\end{equation}
where the coefficients $\alpha$ and $\beta$ can be obtained by diagonalizing
the model Hamiltonian on the tetraquark wave function.
When the two identical heavy quarks are in the $\bar{\textbf{3}}$ color representation and in an $S$-wave, the spin must be equal to $1$ so as to satisfy the Pauli principle. For the two heavy quarks being in the ${\textbf{6}}$ color representation and in an $S$-wave, the spin needs to be 0. Therefore, the mixing between $\bar{\textbf{3}}$ and ${\textbf{6}}$ color representations requires a flip of heavy quark spin, and thus is of higher order in NREFT (suppressed by $v^2\sim0.1$).
Since the OGE gives an attractive interaction for $\bar{\textbf{3}}$ and a repulsive interaction for ${\textbf{6}}$, at LO the ground-state color configuration should be
$\left| \bar{\textbf{3}}_{12}\textbf{3}_{34};
\textbf{1}_{1234} \right\rangle$.
Such a ground-state configuration has been suggested~\cite{Jaffe:2004ph,Santopinto:2006my}, and more recently in Refs.~\cite{Eichten:2017ual,Wu:2016vtq,Park:2013fda}. It is also supported by recent lattice QCD studies on
fully-heavy four quark systems~\cite{Cardoso:2012uka}.

On the basis of the $\left| \bar{\textbf{3}}_{12}\textbf{3}_{34};
\textbf{1}_{1234} \right\rangle$ color configuration, the kinetic
energy matrix elements can be written as
\be
T=\frac{\textbf{p}_{\bm\sigma}^{2}}{2m_{1}}+\frac{\textbf{p}_{\bm\rho}^{2}}{2m_{2}}+\frac{\textbf{p}_{\bm\lambda}^{2}}{2m_{3}}=
\frac{1}{m_b}\bigg(\textbf{p}_{\bm\sigma}^{2}+\textbf{p}_{\bm\rho}^{2}+2\textbf{p}_{\bm\lambda}^{2}\bigg),
\ee
where $m_1$, $m_2$ and $m_3$ are the reduced masses of $Q_1 Q_2$,
$\bar{Q}_3\bar{Q}_4$ and $Q_1 Q_2-\bar{Q}_3\bar{Q}_4$, respectively.
The spatial trial wave function is written in terms of the Jacobi coordinates of
Eq.~(\ref{eqn:Jacobi}) and Fig.~\ref{config},
\be
\psi(\bm\sigma, \bm\rho, \bm\lambda)_{\textrm{spatial}} = \mathcal N \prod_{i=1}^3 \exp\left[-\frac{1}{2} \beta^{2}_i \bm\xi_{i}^{2}\right]  \mbox{ },
\ee
where $(\bm\xi_{1},\bm\xi_{2},\bm\xi_{3})\equiv(\bm\sigma, \bm\rho, \bm\lambda)$,
$\beta_i$ are the oscillatory (variational) parameters, and
\be
\mathcal N= \bigg(\frac{1}{\sqrt{\pi}} \bigg)^{9/2}~\prod_{i} \beta_{i}^{3/2},
\ee
is the overall normalization constant. It has been recently
shown that the above Gaussian variational basis
is a powerful tool to obtain the ground state energy of
heavy tetraquarks~\cite{Park:2013fda}.

It is worth noting that the color configurations  $\left| \textbf{1}_{13} \textbf{1}_{24};
\textbf{1}_{1234} \right\rangle$ and $\left| \textbf{8}_{13} \textbf{8}_{24};
\textbf{1}_{1234} \right\rangle$ are linear superpositions of the two color
representations of Eq.~(\ref{eqn:Psi-c-tetra-phys}). In other words, one can
alternatively use the bases $\left\{ \left| \bar{\textbf{3}}_{12}\textbf{3}_{34};
\textbf{1}_{1234} \right\rangle; \left| \textbf{6}_{12}\bar{\textbf{6}}_{34};
\textbf{1}_{1234} \right\rangle \right\}$ or $\left\{ \left| \textbf{1}_{13}
\textbf{1}_{24}; \textbf{1}_{1234} \right\rangle; \left| \textbf{8}_{13} \textbf{8}_{24};
\textbf{1}_{1234} \right\rangle \right\}$. For the detailed arguments, we refer to Ref.~\cite{Weinstein:1983gd}.
A physical transition between the $\left| \bar{\textbf{3}}_{12}\textbf{3}_{34};
\textbf{1}_{1234} \right\rangle$ tetraquark state and the $\left| \textbf{1}_{13}
\textbf{1}_{24}; \textbf{1}_{1234} \right\rangle$ two-meson state\footnote{Similarly
to the tetraquark model case, for simplicity in the meson-meson molecular model
the color configuration is restricted to $\left| \textbf{1}_{13} \textbf{1}_{24};
\textbf{1}_{1234} \right\rangle$.}
can occur, which is the well-known ``flip-flop'' transition~\cite{Okiharu:2004ve}.
This is an important ingredient while studying the decays into two mesons
if the mass of the initial tetraquark is higher than the two-meson threshold.

\subsection{$b b \bar b \bar b$ ground-state}

The eigenvalue problem is solved by means of a numerical variational
approach with Gaussian trial wave functions~\cite{Hwang:1995uy,Naeem12}.
The quark mass $m_b$ is extracted via Eq.~(\ref{eqn:bmass}), where one
also has to include the spin-averaged mass of the $1S$ (ground-state) bottomonia,
\be
M_{b\bar{b}}(1S)=\frac{1}{4}\left[3M_{\Upsilon(1S)}+M_{\eta_{b}(1S)}\right] = 9.45~\textrm{GeV}
\mbox{ }.
\ee
The other parameter, i.e., the running coupling constant $\alpha_s(\mu)$,
is taken at the scale of the typical momentum transfer in the ground-state bottomonium
($\mu=1.5$~GeV).
The extracted values are
\be
\label{eqn:bb-parameters}
m_b=4.82~\text{GeV} ~, ~~~\alpha_{\rm s}(1.5~{\rm GeV})=0.31\,.
\ee
The spin-averaged binding energy of the $1S$ bottomonium is
\be
\label{eqn:B-bb}
E_{b\bar{b}}(1S) \equiv 2 m_b - M_{b\bar b (1S)} =(0.20\pm0.01)~\textrm{GeV}.
\ee
We can estimate the energy of the $X_{bb\bar{b}\bar{b}}$ ground-state by means of
the variational method with Gaussian trial wave functions previously used in the
bottomonium case and the model parameter values of Eq.~(\ref{eqn:bb-parameters}). We get
\begin{equation}
         \label{eqn:M-bbbb}
	M_{b b \bar b \bar b}^{\rm gs, NR} = (18.72\pm0.02) \mbox{ GeV},
\end{equation}
where the uncertainty is given by the product of the binding energy and $\alpha_{\rm s}(1.5~{\rm GeV})$.
The binding energy of the lowest four $b$-quark bound state is obtained as
\be
\label{eqn:B-bbbb}
E_{bb\bar{b}\bar{b}}(1S) \equiv 4 m_b - M_{b b \bar b \bar b}^{\rm gs, NR}
= (0.56\pm0.02)~\textrm{GeV}.
\ee
The extracted optimal values of the variational parameters $\beta_i$ are
$\beta_1=\beta_2= 0.77$ GeV and $\beta_3= 0.60$ GeV, where $\beta_1=\beta_2$
is due to the symmetry of the spatial wave function~\cite{Hylleraas:1947zza},
as discussed in the last subsection.

\subsection{$b b \bar b \bar b$ ground state in a relativized diquark model}
\label{bbbb relativized}

In this subsection, we provide an estimate of the fully-bottom tetraquark
ground-state energy by treating the $X_{b b \bar b \bar b}$ configuration
as a diquark, $bb$, and antidiquark, $\bar b \bar b$, bound state in a relativized diquark model.
In this model, the diquark is an effective degree of freedom which describes two strongly
correlated quarks with no internal spatial excitations.
The lowest-energy diquark configurations, scalar and axial-vector, are both
color anti-triplets but have different spin and flavor quantum numbers.
The former, {\it i.e.}, the scalar diquark, has spin-0 and its flavor wave
function is antisymmetric. The latter has spin-1 and is flavor-symmetric.
As mentioned in the last section, because of the Pauli principle, a (point-like) diquark made up of quarks
of the same flavor can only be of the axial-vector type \cite{Debastiani:2017msn,Jaffe:2004ph}.

To calculate the energy of the $b b\bar b \bar b$ ground state in diquark anti-diquark configuration, we need an
evaluation of the $b b$ axial-vector diquark mass, $M_{{\rm av}, bb}$.
The diquark (antidiquark) mass is estimated by binding a $bb$ ($\bar b \bar b$)
pair via the OGE plus a confining potential~\cite{Godfrey:1985xj}, and we have\footnote{Notice that the
parameters in this diquark model are different from those in the NREFT
approach.}
\begin{equation}
	M_{{\rm av}, bb} = 9.85~{\rm GeV}.
\end{equation}	
After doing that, we can compute the $1S$, $0^{++}$ tetraquark ground-state
energy in the relativized diquark model. The model Hamiltonian is characterized
by the OGE plus confining potential. For more details on the model and the
values of the model parameters, we refer to Ref.~\cite{Anwar:2018sol}.
With the use of the previous value of $M_{{\rm av}, bb}$, we get
\begin{equation}
	M_{b b \bar b \bar b}^{\rm gs, REL} = 18.75 \mbox{ GeV}\,,
\end{equation}
which is consistent with the result obtained in the previous subsection.
See also Table~\ref{results}, where our relativized diquark model prediction is
compared to those of previous studies.
The theoretical uncertainty in the above prediction can be estimated
by using the typical error of the quark model calculations.
The intrinsic error in the quark model predictions is of the order of $30\sim50$ MeV~\cite{charmonium};
therefore, this diquark model prediction has the cumulative uncertainty of $\mathcal O (50~\textrm{MeV})$.

\section{Results and Discussion}
\label{resultDiss}

In Table~\ref{results}, our NREFT and relativized diquark model results for the
mass of the $b b \bar b \bar b$ ground-state are compared with those of previous
studies.
The results are strongly model-dependent and vary in the range of
$(18.7\pm0.2)$~GeV, approximately.
The differences among the results mainly stem from different choices of effective
Hamiltonians, the model-parameter fitting procedures, and the use of
distinct approximations in the tetraquark wave function.
This last is related to the possible ways of combining the quark color
representations to obtain a color singlet wave function for the tetraquark.
In the last column of the table, we also report the differences between the
predicted masses and the $\eta_b\eta_b$ threshold, $(18798 \pm
4)$~MeV~\cite{Nakamura:2010zzi} whose central value is used.

\begin{table}[h] \renewcommand\arraystretch{2} \centering
\begin{tabular}{lccc}
  \hline\hline
  Reference & Mass [GeV] & $\Delta$ [GeV] \\
  \hline
  This Work (NR) & $18.72\pm0.02$ & $-0.08\pm0.02$\\
  This Work (REL) & $18.75$  & $-0.05$ \\
  Karliner~\textit{et. al.}~\cite{Karliner:2016zzc} & $18.862 \pm 0.025$  &
  $0.06\pm0.03$
  \\
  Bai~\textit{et. al.}~\cite{Bai:2016int}& $18.69\pm 0.03$  & $-0.11\pm0.03$\\
  Berezhnoy~\textit{et. al.}~\cite{Berezhnoy:2011xn} & 18.754  & $-0.04$\\
  Chen~\textit{et. al.}~\cite{Chen:2016jxd} & $18.462 \pm 0.15$  &
  $-0.34\pm0.15$
  \\
  Wu~\textit{et. al.}~\cite{Wu:2016vtq} & $18.462\sim 18.568$ &
  $~-0.28\pm0.05$\\
  Wang~\cite{Wang:2017jtz} & $18.84 \pm 0.09$  & $0.04\pm0.09$\\
  \hline\hline
\end{tabular}
\caption{Different theoretical predictions of masses for the ground state
($J^{PC}=0^{++}$) $X_{bb\bar{b}\bar{b}}$. The acronyms NR and REL refer to
nonrelativistic and relativized results, respectively. In the last column,
$\Delta\equiv m(X_{bb\bar{b}\bar{b}})-2m(\eta_b)$.}
\label{results}
\end{table}

It is worthwhile to remind some previous studies.
For example, Karliner~\textit{et al.}~\cite{Karliner:2016zzc}  made a
phenomenological estimation of the $Q Q \bar Q \bar Q$ ground-state energies,
with $Q = b$ and $c$.
The authors utilized a relation between meson and baryon masses to extrapolate
the binding energy of $Q Q \bar Q \bar Q$ systems in the diquark model.
They made the ballpark estimates of $B_{b\bar{b}}(1S)$ and the binding energy of
the tetraquark with respect to the two-$bb$-diquark threshold, and obtained a
value of about $1/2$ for the ratio ${B_{b\bar{b}}(1S)}/{B_{bb\bar{b}\bar{b}}(1S)}$.

Bai~\textit{et al.} calculated the $b b \bar b \bar b$ ground-state energy by
means of a phenomenological potential, whose parameters were fitted to the $b
\bar b$ spectrum and verified by lattice simulation~\cite{Bai:2016int}. The $b b
\bar b \bar b$ ground-state energy was obtained by solving the Schr\"odinger
equation numerically; spin-dependent corrections and a linear confining
potential were included.
Their result agrees with our LO NREFT result very well, indicating that
neglecting the spin-dependent and confining parts provides a very good
approximation to the system under study.

A diquark model prediction of the  $b b \bar b \bar b$ and $c c \bar c \bar c$
ground-state energies was given in Ref.~\cite{Berezhnoy:2011xn}.
There, the authors assumed tetraquarks to be made up of two almost point-like
diquarks in the color-triplet configuration. The model parameters were
fitted by solving a non-relativistic Schr\"odinger equation for the charmonium
and bottomonium spectrum. The $bb\bar b\bar b$ ground-state mass agrees with our
diquark model well.

In Ref.~\cite{Ader:1981db}, the authors computed the $c c \bar c \bar c$
ground-state energy within a nonrelativistic potential model.
The authors concluded that the possibility of obtaining bound states depends on
the assumptions made on the quark dynamics and the flavor configurations. There
are also early predictions for the mass~\cite{Iwasaki:1975pv} and width~\cite{chao:1981} of the
fully-charm tetraquark, and the ground-state energies of $Q^2 \bar{Q}^2$ systems, with
$Q\in \{s,c,b,t\}$~\cite{heller:1985}.
For the full mass spectrum of all-charm tetraquarks, we refer to Ref.~\cite{Debastiani:2017msn}.
All-charm four-quark bound state has also been studied in the Bethe--Salpeter approach and
the ground state was reported to be deeply bound (650 MeV) below the
$2\eta_c$ threshold~\cite{Heupel:2012ua}.

A study of the importance of mixing effects
between tetraquark and molecular-type components is also worthwhile to be
carried out, especially for those states which are close to meson-meson
thresholds. The probability of a fully-heavy tetraquark to be a
$(Q\bar{Q})-(Q\bar{Q})$ hadronic molecule might be calculable without any
ambiguity. It is worthwhile to mention that the possibility of an
$\eta_{\rm b}$-$\eta_{\rm b}$ bound state has been studied by computing the QCD van der
Vaals force in the framework of potential nonrelativistic QCD in Ref.~\cite{Brambilla:2016}.
Given that the recent lattice
nonrelativistic QCD calculation did not find any state below the noninteracting
$\eta_b\eta_b$ threshold~\cite{Hughes:2017xie}, such a study seems necessary to
understand what happens.

It would also be very interesting to test the possibility of a $bb\bar c\bar c$
tetraquark that remains stable against strong decays,\footnote{All the other
possible fully-heavy tetraquarks can decay strongly by annihilating at least
a pair of quarks and antiquarks of the same flavor.} but unfortunately
there is no experimental evidence yet.
The stability of heavy-light tetraquark systems is still an open question.
The $QQ \bar{q} \bar{q}$ was shown to be stable against strong decays by Lipkin long ago\footnote{Before Lipkin,
the authors of Ref.~\cite{Ader:1981db} also argued
that $X_{QQ \bar{q} \bar{q}}$ is stable if the quark mass ratio $m_Q/m_q$ is large enough.
For reasonable values of the quark mass ratio, $5\sim20$, they predict a
$X_{QQ \bar{q} \bar{q}}$ above but not very far from the corresponding threshold.
Note that the ratios $m_b/m_u\approx 16$, $m_c/m_u\approx 5$, and $m_b/m_c\approx 3$,
if one uses typical values for the constituent quark masses.}~\cite{lipkin85}.
Very recently, 
$bb \bar{q} \bar{q}$ was shown to be stable against strong decays but not its charm
counterpart $cc \bar{q} \bar{q}$, nor the mixed (beauty+charm) $bc \bar{q} \bar{q}$
state~\cite{Karliner:2017qjm,Eichten:2017ffp}.
For a detailed discussion on the stability of such systems, we refer to the
recent study in Ref.~\cite{Czarnecki:2017vco}.

\section{Tetraquark-Meson Mass Inequalities}
\label{massInqSec}

One can estimate the lower bounds of the ground state energy levels of the
fully-heavy four-quark systems by using the variational approach, as suggested
long ago by Nussinov~\cite{Nussinov:1983vh} and by Bertlmann and Martin~\cite{Bertlmann:1979zs}.
This approach was tested by the lattice QCD calculations of
Weingarten~\cite{Weingarten:1983uj} and also by the rigorous calculations in the
vectorlike gauge theories (QCD) by Witten~\cite{Witten:1983ut}.
We attempt to extend Nussinov's approach to the fully-heavy four-quark bound
states and obtain inequalities between the tetraquark states and the
corresponding heavy
quarkonia in the following.

Let us consider the general Hamiltonian of four heavy quarks with pair-wise
interactions:
\begin{equation}
   \begin{array}{rcl}
H_{4}(Q_1 Q_2 \bar{Q}_3\bar{Q}_4) &=& \sum_{i=1}^{4} T_i (Q_i)+ V_{Q_1 Q_2} + V_{\bar{Q}_3 \bar{Q}_4} \\
&& + V_{Q_1 \bar{Q}_3}+ V_{Q_1 \bar{Q}_4}+ V_{Q_2 \bar{Q}_3}+ V_{Q_2 \bar{Q}_4}
\,,
\end{array}
\label{H4q}
\end{equation}
The color-antitriplet potential $V_{QQ}^{(\bar{\textbf{3}})}$ between any quark
(or antiquark) pair
can be related to the color-singlet quark-antiquark potential,
$V_{Q\bar{Q}}^{(\textbf{1})}$~\cite{Iwao:1984pr}, via
\be
V_{QQ}^{(\bar{\textbf{3}})}=\frac{1}{[N_c-1]} V_{Q\bar{Q}}^{(\textbf{1})},
\label{eq:vqqineq}
\ee
where $1/[N_c -1]$ is the overall ratio of the color coefficient in the
leading SU($N_c$) group. The above relation can also be verified by using
the eigenvalues of the Casimir invariants from the Table~\ref{casimir},
and it holds when the $QQ$ and $Q\bar{Q}$
pairs are in the same spin and orbital quantum states. With the use of
$V_{QQ}^{(\bar{\textbf{3}})}=\dfrac{1}{2}V_{Q\bar{Q}}^{(\textbf{1})}$ for SU(3),
Eq.~(\ref{H4q})
can be rearranged as follows
\begin{eqnarray}
H_{4}(Q_1 Q_2 \bar{Q}_3\bar{Q}_4) &=& \sum_{i=1}^{4} T_i (Q_i) + \dfrac{1}{2}
\bigg (
V_{12}^{(\textbf{1})} + V_{34}^{(\textbf{1})}
\bigg) \nonumber\\
&& +\sum_{i=1,2;j=3,4} V_{Q_i \bar{Q}_j}\mbox{ },
\label{H4q2}
\end{eqnarray}
where as before we consider only the color-antitriplet for $Q_iQ_j$ and
triplet for $\bar Q_i\bar Q_j$, and we have written the
color-singlet quark-antiquark ($Q_i\bar Q_j$)
potential from applying Eq.~\eqref{eq:vqqineq} to $Q_iQ_j$ as
$V_{ij}^{(\textbf{1})}$.

\begin{table}
  \renewcommand\arraystretch{2.2}
  \centering
\begin{tabular}{p{3cm} c | p{3.5cm} c}
  \hline\hline
  $\big|\Psi_{ij} \big\rangle_{\textrm{color}}$ & $\hat{C}_i \cdot \hat{C}_j$ &
$\big|\Psi\big\rangle_{\textrm{color}}$ & $\hat{C}^2$\\ \hline
  $\big|Q_i \bar{Q}_j \big\rangle_\mathbf{1}$ &  $-\dfrac{4}{3}$ &
$\big|\Psi\big\rangle_{\textrm{singlet}}$ & $0$ \\
  $\big|Q_i \bar{Q}_j \big\rangle_\mathbf{8}$ &  $+\dfrac{1}{6}$ &
$\big|\Psi\big\rangle_{\textrm{octet}}$ & $3$ \\
  $\big|Q_i Q_j \big\rangle_\mathbf{\bar{3}}$ & $-\dfrac{2}{3}$ &
$\big|\Psi\big\rangle_{\textrm{triplet/antipriplet}}$ & $\dfrac{4}{3}$ \\
  $\big|Q_i Q_j \big\rangle_\mathbf{6}$ & $+\dfrac{1}{3}$  &
$\big|\Psi\big\rangle_{\textrm{sextet}}$ & $\dfrac{10}{3}$ \\
  \hline\hline
\end{tabular}
\caption{Eigenvalues of the Casimir invariants $ \hat{C}_i \cdot
\hat{C}_j$ and $\hat{C}^2$.
}
\label{casimir}
\end{table}

To work out the last term in the above equation,
one has to calculate the eigenvalues of the color matrices, viz.
the Casimir invariants
\be
\sum_{i=1,2;j=3,4} V_{Q_i \bar{Q}_j} \propto \Big\langle \bar{\textbf{3}}_{12}
\textbf{3}_{34} \Big| \sum_{i,j} \hat{C}_i \cdot \hat{C}_j \Big|
\bar{\textbf{3}}_{12} \textbf{3}_{34} \Big\rangle~,
\ee
where $\hat{C}_{i,j}= \vec{\lambda}_{i,j}/2$. The eigenvalues of these Casimir
invariants are explicitly given in Table~\ref{casimir}. We need to know in
which
color representation the $Q_i \bar Q_j$ pair is.
For instance, by writing down
explicitly the color wave function (see, e.g., Ref.~\cite{Stancu}), we can get
\begin{eqnarray}
\big| \bar{\textbf{3}}_{12} \textbf{3}_{34} \big\rangle_\textbf{1} &=&
 \frac1{\sqrt{3}}  \big| \textbf{1}_{13} \textbf{1}_{24}\big\rangle -
\sqrt{\frac23} \big| \textbf{8}_{13} \textbf{8}_{24} \big\rangle  \nonumber\\
 &=&  -\frac1{\sqrt{3}}  \big| \textbf{1}_{14} \textbf{1}_{23}\big\rangle +
\sqrt{\frac23} \big| \textbf{8}_{14} \textbf{8}_{23} \big\rangle\,.
\end{eqnarray}
Using $\hat{C}_i \cdot \hat{C}_j = \left(\hat C_{ij}^2 - \hat C_i^2 - \hat
C_j^2 \right)/2$, where $\hat C_{ij} = \hat C_i + \hat C_j$, we have
\begin{eqnarray}
\big\langle \bar{\textbf{3}}_{12} \textbf{3}_{34} \big| \hat{C}_{i} \cdot
\hat{C}_{j} \big| \bar{\textbf{3}}_{12} \textbf{3}_{34} \big\rangle &=&
 \frac{1}{4}~\big\langle \bar{\textbf{3}}_{12} \textbf{3}_{34} \big|
\hat{C}_{12} \cdot \hat{C}_{34} \big| \bar{\textbf{3}}_{12} \textbf{3}_{34}
\big\rangle \nonumber\\
&=& - \frac13
\end{eqnarray}
for $i=1,2; j=3,4$.
This result enables us to write the pair-wise potential
of four-quark state in terms of the quark-antiquark color-singlet potential, viz.,
the quarkonium potential,
\be
\sum_{i<j} V_{ij} =
\frac{1}{2} \bigg (V_{12}^{(\textbf{1})} + V_{34}^{(\textbf{1})} \bigg)+
\frac{1}{4} \bigg( V_{13}^{(\textbf{1})} + V_{14}^{(\textbf{1})} +
V_{23}^{(\textbf{1})} + V_{24}^{(\textbf{1})} \bigg)\,.
\ee

This means that under the approximation of one-gluon exchange and that the two
quarks are in color anti-triplet, the four-quark Hamiltonian [(Eq.~(\ref{H4q2})] can
be
expressed in the following form,
\begin{eqnarray}
H_{4}(Q_1 Q_2 \bar{Q}_3\bar{Q}_4) &=& \frac{1}{2} \Big( H_{12} + H_{34} \Big) \nonumber\\
&& +
\frac{1}{4} \Big( H_{13} + H_{14} + H_{23}  + H_{24} \Big),~~~
\end{eqnarray}
where $H_{ij}= T_i + T_j + V_{ij}^{(\textbf{1})} (\textbf{r}_{ij})$ is the
quarkonium Hamiltonian. Taking quarkonium wave functions as the trial wave
function, and applying the variational principle~\cite{Nussinov:1983vh},
we obtain an upper bound on the ground state energy of four-heavy-quark
system
by computing the expectation value of the Hamiltonian of the subsystems, namely
\begin{eqnarray}
E_{4Q} &\equiv&  \big\langle H_{4}(Q_1 Q_2 \bar{Q}_3\bar{Q}_4) \big\rangle \nonumber\\
&\lesssim &
\frac12\Big(\big\langle \psi_{12} \big| H_{12} \big|\psi_{12} \big\rangle +
\big\langle \psi_{34} \big| H_{34} \big| \psi_{34} \big\rangle\Big) \nonumber\\
&& + \frac14
\sum_{i=1,2;j=3,4} \big\langle \psi_{ij} \big| H_{ij} \big| \psi_{ij} \big\rangle \,,
\label{E4Q}
\end{eqnarray}
where $\psi_{ij}$ are the corresponding ground state wave functions of the $ij$
subsystems. Here, we use $\lesssim$ instead of $\leq$ because we have made the approximation that the two quarks are in color anti-triplet and the two anti-quarks are in color triplet, though it is expected to work rather well since the mixing with the sextet-antisextet configuration is expected to be suppressed by $v^2\sim0.1$ for fully-bottom and $\sim0.3$ for fully-charm four-quark systems. In this sense, the inequalities derived for baryons by Nussinov~\cite{Nussinov:1983vh} are more rigorous because the two quarks in a baryon must be in an anti-triplet without any approximation.

\begin{table*}
  \renewcommand\arraystretch{2}
  \centering
   \begin{tabular}{lcc}
  \hline\hline
  State  & Mass Inequality & Upper Bound (GeV) \\
  \hline
  $X_{bb\bar{b}\bar{b}}$ & $\lesssim 2 M_{b\bar{b}}(1S)  $ & $\lesssim 18.89$ \\
  $X_{bb\bar{b}\bar{c}}$ & $\lesssim M_{b\bar{b}}(1S) + M_{c\bar{b}}(1S)$ & $\lesssim 15.77$ \\
  $X_{b\bar{b}c\bar{c}}$ & $\lesssim \dfrac{1}{4} \bigg(M_{b\bar{b}}(1S)+M_{c\bar{c}}(1S)\bigg)+ \dfrac{3}{2} M_{c\bar{b}}(1S)$ & $\lesssim 12.62$ \\
  $X_{bb\bar{c}\bar{c}}/X_{cc\bar{b}\bar{b}}$ & $\lesssim \dfrac{1}{2} \bigg(M_{b\bar{b}}(1S)+M_{c\bar{c}}(1S)\bigg)+M_{c\bar{b}}(1S)$ & $\lesssim 12.58$ \\
  $X_{cc\bar{c}\bar{b}}$ & $\lesssim M_{c\bar{c}}(1S) + M_{c\bar{b}}(1S)$ & $\lesssim 9.39$ \\
  $X_{cc\bar{c}\bar{c}}$ & $\lesssim 2 M_{c\bar{c}}(1S)  $ & $\lesssim 6.14 $ \\
  \hline\hline
\end{tabular}
\caption{Mass inequality relations and upper bounds of the mass of ground-state fully-heavy tetraquarks formed of bottom and charm quarks.}
\label{massInq}
\end{table*}

The inequalities for all possible fully-heavy tetraquarks are listed in  Table~\ref{massInq}. The corresponding numerical values are obtained using the spin-averaged quarkonium masses $M_{b\bar{b}}(1S)=9.445~\text{GeV}$, $M_{c\bar{c}}(1S)= 3.069~\text{GeV}$ and $M_{c\bar{b}}(1S)= 6.324~\text{GeV}$. Since the $B_{c}^* (1 ^3S_1)$ meson has not been observed, for $M_{c\bar{b}}(1S)$ we used the average value of the theoretical predictions of
Godfrey--Isgur Model~\cite{Godfrey:2004ya} and the Cornell potential model
from Ref.~\cite{Eichten:1994gt}.

If we compare the so-obtained approximate upper bound in the fully-bottom sector with the
earlier theoretical predictions listed in Table~\ref{results}, we find this
value is indeed larger than all the model predictions except for that of
Ref.~\cite{Wang:2017jtz}, which is calculated by means of QCD sum rules and has a large
uncertainty.

\section{Possible Decays of $X_{bb\bar{b}\bar{b}}$}
\label{decays}

The main difficulties in the experimental observation of fully-bottom
tetraquark mesons are related to the production mechanisms and observation
of the main decay modes. A very recent study discussed the
production of the $bb\bar b\bar b$ ground-state at the LHC,
concluding that the $bb\bar b\bar b$ is supposed to be very narrow and
likely to be discovered \cite{Eichten:2017ual}.
Another recent study~\cite{Vega-Morales:2017pmm} also discussed
the possible production of a narrow scalar resonance around $18\sim19$ GeV
at the LHC.\footnote{The decay of such resonance into four-lepton final state is
also explored to disentangle whether it would be a tetraquark or something more exotic.}
In the following, we will discuss the decays of such a state  and provide a rough
estimate of its width. We will argue that the width is at least a few tens of MeV.

\begin{figure}[tb]
\includegraphics[width=5.5cm]{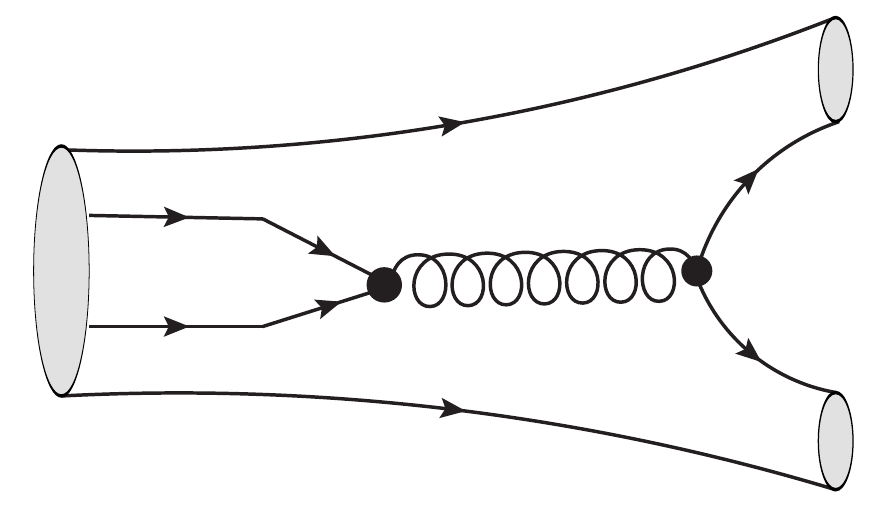}
\caption{Quark level description of hadronic decays $X_{bb\bar{b}\bar{b}}  \to
M_1 \bar{M}_2$, where $M_1$ and $\bar{M}_2$ are spin-parity and phase-space
allowed bottom- and anti-bottom meson, respectively.}
\label{dec}
\end{figure}

As shown in Table~\ref{results}, most of the models predict the ground-state
fully-bottom tetraquark to be below the $\eta_b\eta_b$ threshold. Because of this,
one expects its width to be almost saturated by the following decay modes:
1) decays into final states containing a pair of bottom and anti-bottom quarks;
2) decays into hadrons of lighter flavors. The former is dominated by one-gluon exchange,
while the latter should be dominated by two gluons. Figure~\ref{dec} shows an
example of the decay into a pair of open-bottom mesons.
Because the gluons are at the scale of $m_b$, one expects the first type
of decays to dominate over the second. As it is not an easy task to calculate
partial decay widths into given exclusive decay modes, in the following we
provide a rough estimate of the inclusive decay width of the
$X_{bb\bar b\bar b}$ ground-state based on the first decay mode.

The inclusive width of the $X_{bb\bar{b}\bar{b}}$ into final states with a pair
of $b$ and $\bar b$ quarks can be described as a two-step process triggered by a
transition operator $\mathcal T$ as
\begin{equation}
\begin{array}{rcl}
\langle h_1 h_2\ldots | \mathcal T| X_{bb\bar{b}\bar{b}}\rangle & = & \langle h_1
h_2... | \mathcal T_2| b_1\bar{b}_4 g\rangle \\
& & \cdot \langle b_1\bar{b}_4 g | \mathcal T_1| b_1 b_2 \bar{b}_3\bar{b}_4
\rangle\,,
\end{array}
\label{amp}
\end{equation}
where $h_1$ and $h_2$ indicate possible bottom hadrons allowed by the
spin-parity quantum numbers and the available phase space. Here
$ \mathcal T_1$ is operator for the coupling of a heavy quark-antiquark pair to a
gluon, given by
$\sqrt{4\pi \alpha_s}~\bar{Q} \frac{1}{2} \lambda_a \gamma^{\mu} Q
\epsilon^{a}_{\mu}$, and $\mathcal T_2$ is responsible for the transition from
$b\bar b g $ to the final hadronic states.
We need to sum up all the possible final states generated at
the second vertex of Fig.~\ref{dec}, which includes not only two-body but also many-body final states.
As a result, only the second factor ($\mathcal T_1$) in the above equation matters.

The $\bar{Q}Q$ annihilation at short distances requires
a direct dependence on the zero-point  wave function of the color-octet $\bar Q Q$
inside the four-body bound state, $R_{Q\bar{Q}_{(8)}}(0)$.
Hence,
\be
\Gamma(X_{bb\bar{b}\bar{b}}\to h_1h_2\ldots) \propto
\alpha_s(m_b)~|R_{b\bar{b}_{(8)}}(0)|^2 \mbox{ }.
\ee
It is well-known that the $\eta_b$ decay width is saturated
by two-gluon exchange, and the $\eta_b$ inclusive decay width is
$\Gamma(\eta_{b} \to \mbox{hadrons}) \propto \alpha_{s}^2(m_b)~
|R_{b\bar{b}_{(1)}}(0)|^2$~\cite{voloshin:1978Rosner:1988}.
The wave function at origin of a color-octet
$b\bar{b}_{(8)}$
pair might be larger than that of the asymptotic color-singlet $b\bar{b}_{(1)}$,
namely a bottomonium~\cite{Isgur:1984bm}.
However, at the present stage there is no need for a precise calculation of the
width. For an order-of-magnitude estimate, one may simply assume
$|R_{b\bar{b}_{(8)}}(0)|^2 \sim |R_{b\bar{b}_{(1)}}(0)|^2$.
Therefore, we can estimate the inclusive width of the $X_{bb\bar{b}\bar{b}}$
decays into hadrons containing $b$ and $\bar b$ as
\be
\Gamma(X_{bb\bar{b}\bar{b}}\to h_1h_2\ldots) \simeq \frac{1}{\alpha_s(m_b)}
\Gamma (\eta_{b} \to \mbox{hadrons}) \mbox{ }.
\ee
The $\eta_{\rm b}$ dominantly decays into hadrons, hence, $\Gamma (\eta_{b}
\to \mbox{hadrons})\approx
\Gamma(\eta_{b})=10^{+5}_{-4}$~MeV~\cite{Nakamura:2010zzi}. Taking
$\alpha_s(m_b)=0.22$~\cite{Nakamura:2010zzi}, and neglecting all other possible
decay modes which should be subdominant, we get
\be
\label{eqn:bbbb-width}
\Gamma(X_{bb\bar{b}\bar{b}}) =\mathcal{O}(50~\textrm{MeV})\, .
\ee
The previous width (of the order of a few tens of MeV) is large enough
to make the resonance observable. Our estimate of Eq. (\ref{eqn:bbbb-width})
is of the same order of magnitude as an earlier prediction for similar systems~\cite{chao:1981},
while it is much larger than the estimate of Ref.~\cite{Karliner:2016zzc}, 1.2~MeV.
Similarly, we expect the width of the fully-charm tetraquark, if below the $\eta_c\eta_c$ threshold, to be larger,
\begin{equation}
  \Gamma(X_{cc\bar c\bar c}) \simeq \frac{\Gamma(\eta_c)}{\alpha_s(m_c)} = \mathcal{O}(100~\textrm{MeV})\,.
\end{equation}

The fully-bottom tetraquark states could be searched for in final states
including a pair of bottom hadrons, such as $B\bar B$, $\Lambda_b\bar
\Lambda_b$, $\Xi_b\bar \Xi_b$, $\Sigma_b\bar \Sigma_b$ and $\Omega_b\bar \Omega_b$.
They can also decay into a fully leptonic final state via an intermediate
$\Upsilon(1S)X$ state as
\be
\label{eqn:Xinto4l}
X_{bb\bar{b}\bar{b}} \to \Upsilon(1S)X \to l^+l^- l^+l^-,
\ee
where $l$ can be $\tau$, $\mu$ or $e$, and $X$ could be the off-shell lowest
vector bottomonium, $X\equiv\Upsilon(1S)^*$.
This decay involves the annihilation of two $b\bar{b}$ pairs into virtual
photons, so the branching fraction is
expected to be small, $\mathcal O (10^{-4}\sim 10^{-8})$, as estimated
in~\cite{Karliner:2016zzc}.\footnote{We have down-scaled the branching faction
estimate by one order of magnitude in comparison with that in
Ref.~\cite{Karliner:2016zzc}, since the total width estimated here is one order
of magnitude larger.} Despite of this, the multi-lepton final states are expected
to provide a clean signal with a low background.
The ideal place to look for the decays of Eq.~(\ref{eqn:Xinto4l}) is the Large
Hadron Collider experiments, where the Higgs boson cross section was measured by
reconstructing a four-lepton final state~\cite{cms:2015yvw}.
Because of this, a scan at relatively lower energies, of the order of
$2M_{\eta_{b}(1S)}$, should be almost straightforward
at LHC.

\section{Summary}

We calculated the $b b \bar b \bar b$ ground-state energy in terms of
two different approximations for the tetraquark wave function.
They were used to simplify the solution of the four-body problem.

In the first case, we provided an evaluation of the $b b \bar b \bar b$
ground-state energy in an NREFT at the LO, where the potential is an OGE-induced color Coulomb potential.
A nice feature of our approach is that the color wave function is completely given by $\left| \bar{\textbf{3}}_{12}\textbf{3}_{34};
\textbf{1}_{1234} \right\rangle$ at the LO, and its mixing with $\left| \textbf{6}_{12}\bar{\textbf{6}}_{34};
\textbf{1}_{1234} \right\rangle$ only occurs at higher orders.
The solution of the four-body problem was simplified by making use of the
symmetries introduced by Hylleraas and Ore in their study of polyelectrons,
namely the bound states of two electrons and positrons.
In our specific case, one of the above mentioned symmetries could be expressed as
$\psi_{\textrm{spatial}}(bb)=\psi_{\textrm{spatial}}(\bar b \bar b)$, where
$\psi_{\textrm{spatial}}(bb)$ and $\psi_{\textrm{spatial}}(\bar b \bar b)$ are
the spatial wave functions of the $bb$ and $\bar b \bar b$ systems,
respectively.
Thanks to this, the four-body problem was simplified by reducing the number
of integration variables. In the second case, we calculated the $b b \bar b \bar b$
ground-state energy in a relativized diquark model. This model is characterized by
OGE plus a confining potential. In the diquark model, the effective degree of
freedom of the diquark, describing two strongly correlated quarks with no
internal spatial excitations, is introduced. Tetraquark mesons are then
obtained as two-body diquark-antidiquark bound states.
Our results in both approaches---NREFT at LO ($18.72\pm0.02$~GeV) and relativized diquark model (18.75~GeV)---only differ by a few tens of MeV, and suggests the existence of a $b b \bar b \bar b$ bound-state below the $\eta_b\eta_b$ threshold.

We also derived a set of approximate inequalities for the binding energies of the
fully-heavy tetraquarks in terms of those of various heavy quarkonia.
Instead of giving the values of the ground-state energies of the
states of interest, the inequalities provide upper bounds on them.
As expected, our LO NREFT and relativized diquark model results on the $X_{bb\bar{b}\bar{b}}$ ground-state mass satisfy the corresponding inequality.

Finally, we discussed the possible decay modes of the lowest fully-heavy
tetraquarks and estimated the decay width of the ground state
$bb\bar b\bar b$ to be of $\mathcal{O}(50~\text{MeV})$.
We hope our results might provide useful references in the
search for fully-heavy tetraquarks in ongoing LHC experiments.
In particular, we suggest searching for the  lowest $X_{bb\bar{b}\bar{b}}$
in the relevant center-of-mass energy region around 18.7~GeV in the
final states of four leptons or a pair of bottom hadrons  such as $B\bar B$,
$\Lambda_b\bar \Lambda_b$, $\Xi_b\bar \Xi_b$, $\Sigma_b\bar \Sigma_b$ and
$\Omega_b\bar \Omega_b$, \blue{  }

\section*{Note added}

During the revision process of this manuscript, a preliminary CMS analysis
(not yet approved by the collaboration) was presented in APS April
meeting~\cite{Durgut},
hinting at a potential four-lepton excess in the mass range
of $18\sim19$~GeV.

\begin{acknowledgments}

The authors are grateful to Ahmed Ali, M. Ahmad, M. Jamil Aslam, Chirstoph Hanhart, Yu Lu and J.-M. Richard
for useful discussions and suggestions. M.N.A. is indebted to Ulf-G.~Mei{\ss}ner for the hospitality
provided at the Helmholtz-Institut f\"ur Strahlen- und Kernphysik (HISKP) at
Universit\"at Bonn, where part of this work was carried out.
This work is supported in part by the National Natural Science Foundation of
China (NSFC) under Grant No.~11621131001 (CRC110 by DFG and NSFC), and Grant
No.~11747601, by the Thousand Talents Plan for Young Professionals, by the
CAS Key Research Program of Frontier Sciences (Grant No. QYZDB-SSW-SYS013), and by the CAS Center for Excellence in Particle Physics (CCEPP).
M.N.A. also received support from the CAS-TWAS President's Fellowship for International
Ph.D. Students.

\end{acknowledgments}

\end{document}